\documentclass{ws-procs975x65standard}

\let\trimmarks\relax

\begin{document}

\title {High-Energy Neutrino Astronomy\footnotemark}
\author{\vskip-.2in Francis Halzen}
\address{Department of Physics, University of Wisconsin, Madison, WI 53706}

\maketitle

\footnotetext{Talk presented at Thinking, Observing and Mining the Universe, Sorrento, Italy, September 2003.}

\vskip-1.75in
\hbox to \hsize{{\bf University of Wisconsin - Madison}\hfill
\vtop{\hbox{\bf MADPH-03-1362}\hbox{November 2003}}}
\vskip1.35in

\abstracts{
Kilometer-scale neutrino detectors such as IceCube are discovery instruments covering nuclear and particle physics, cosmology and astronomy. Examples of their multidisciplinary missions include the search for the particle nature of dark matter and for additional small dimensions of space. In the end, their conceptual design is very much anchored to the observational fact that Nature accelerates protons and photons to energies in excess of $10^{20}$ and $10^{13}$\,eV, respectively. The cosmic ray connection sets the scale of cosmic neutrino fluxes. In this context, we discuss the first results of the completed AMANDA detector and the reach of its extension, IceCube. Similar experiments are under construction in the Mediterranean. Neutrino astronomy is also expanding in new directions with efforts to detect air showers, acoustic and radio signals initiated by super-EeV neutrinos.}

\section{Neutrinos Associated with the Highest Energy Cosmic Rays}

The flux of cosmic rays is summarized in Fig.\,1a,b\cite{gaisseramsterdam}. The  energy spectrum follows a broken power law. The two power laws are separated by a feature referred to as the ``knee"; see Fig.\,1a. There is evidence that cosmic rays, up to several EeV, originate in galactic sources. This correlation disappears in the vicinity of a second feature in the spectrum dubbed the ``ankle". Above the ankle, the gyroradius of a proton exceeds the size of the galaxy and it is generally assumed that we are  witnessing the onset of an extragalactic component in the spectrum that extends to energies beyond 100\,EeV. Experiments indicate that the highest energy cosmic rays are predominantly protons. Above a threshold of 50 EeV these protons interact with CMBR photons and therefore lose their energy to pions before reaching our detectors. This limits their sources to tens of Mpc, the so-called Greissen-Zatsepin-Kuzmin cutoff.

\begin{figure}[ht]
\centering\leavevmode
\epsfxsize=5in
\epsffile{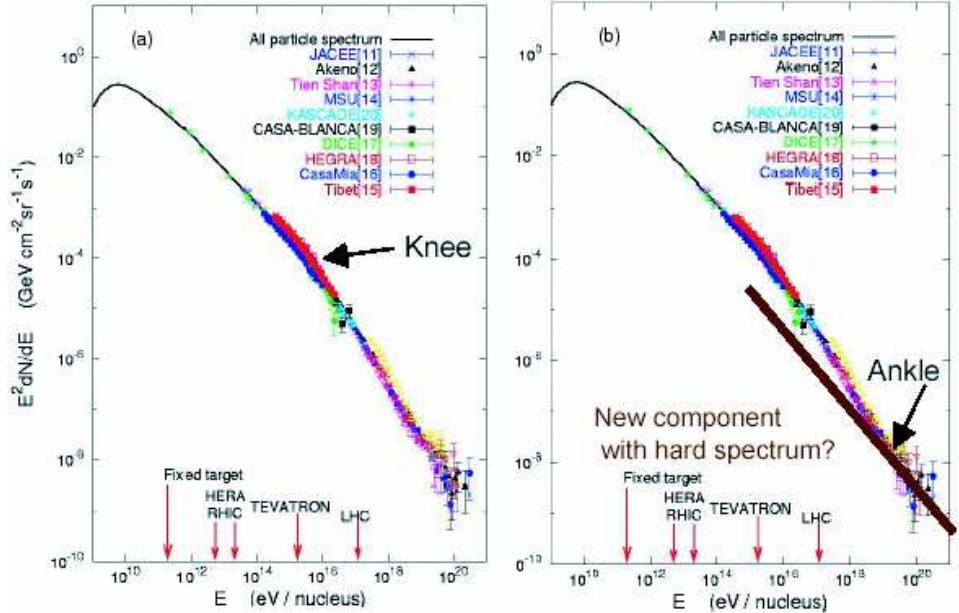}
\caption[]{At the energies of interest here, the cosmic ray spectrum consists of a sequence of 3 power laws. The first two are separated by the ``knee" (left panel), the second and third by the ``ankle". There is evidence that the cosmic rays beyond the ankle are a new population of particles produced in extragalactic sources; see right panel.}
\end{figure}

Models for the origin of the highest energy cosmic rays fall into two categories, top-down and bottom-up. In top-down models it is assumed that the cosmic rays are the decay products of cosmological remnants with Grand Unified energy scale $M_{GUT} \sim 10^{24}\rm\,eV$. These models predict neutrino fluxes most likely within reach of AMANDA, and certainly IceCube.

In bottom-up scenarios it is assumed that cosmic rays originate in cosmic accelerators. Accelerating particles to TeV energy and above requires massive bulk flows of relativistic charged particles. These are likely to originate from the exceptional gravitational forces  in the vicinity of black holes. Examples include the dense cores of exploding stars, inflows onto supermassive black holes at the centers of active galaxies and annihilating black holes or neutron stars. Before leaving the source, accelerated particles pass through intense radiation fields or dense clouds of gas surrounding the black hole. This results in interactions producing pions decaying into secondary photons and neutrinos that accompany the primary cosmic ray beam as illustrated in Fig.\,2.

\begin{figure}[ht]
\centering\leavevmode
\epsfxsize=4.5in
\epsffile{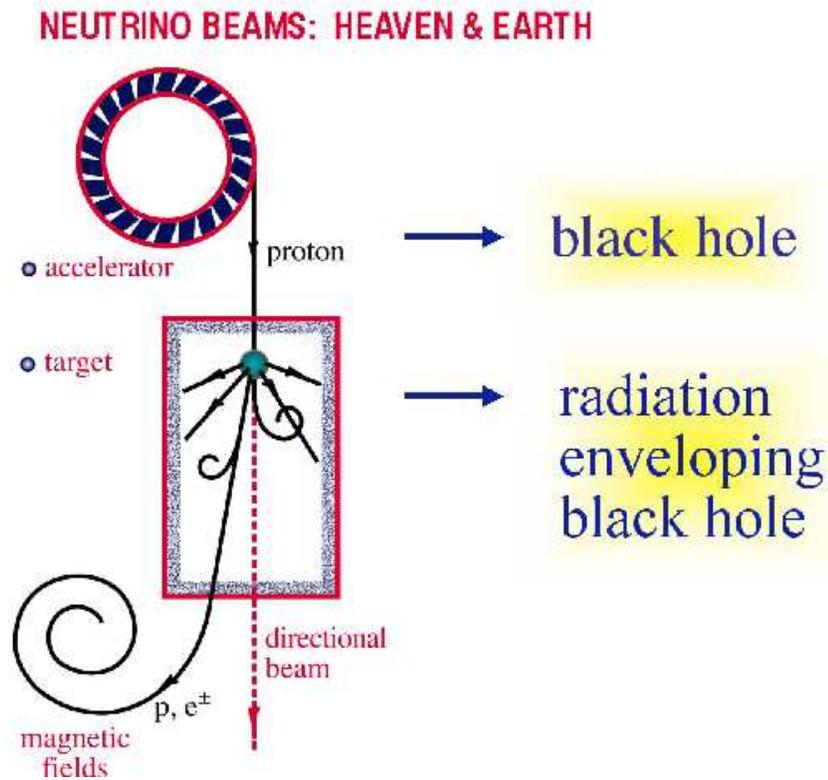}
\caption{Diagram of cosmic ray accelerator producing photons and neutrinos.}
\end{figure}

How many neutrinos are produced in association with the cosmic ray beam? The answer to this question, among many others\cite{PR},  provides the rationale for building kilometer-scale neutrino detectors. We first consider a neutrino beam produced at an accelerator laboratory; see Fig.\,2. Here the target absorbs all parent protons as well as the muons, electrons and gamma rays produced. A pure neutrino beam exits the dump. If nature constructed such a ``hidden source" in the heavens, conventional astronomy will not reveal it. It cannot be the source of the cosmic rays, however, because the dump would have to be partially transparent to protons. The extreme opposite case is a ``transparent source" where the accelerated proton interacts once and escapes the dump after producing photons as well as neutrinos. Elementary particle physics is now sufficient to relate all particle fluxes because a fraction (1/6 to 1/2 depending on the energy) of the interacting proton goes into pion production. This energy is equally shared between gamma rays and neutrinos, of which one half are muon-neutrinos. Therefore, at most one quarter of the energy ends up in muon-neutrinos compared to cosmic rays. The flux of a transparent cosmic ray source is often referred to as the Waxman-Bahcall flux\cite{wb1}. It is easy to derive and the derivation is revealing.

Fig.\,1b shows a fit to the observed cosmic ray spectrum assuming an extragalactic component fitted above the ``ankle". The energy content of this component is $\sim 3 \times 10^{-19}\rm\,erg\ cm^{-3}$, assuming an $E^{-2}$ energy spectrum with a GZK cutoff. The power required to generate this energy density in the Hubble time of $10^{10}$\,years is $\sim 3 \times 10^{37}\rm\,erg\ s^{-1}$ per (Mpc)$^3$. This works out to\cite{TKG}
\begin{itemize}
\item $\sim 3 \times 10^{39}\rm\,erg\ s^{-1}$ per galaxy,
\item $\sim 3 \times 10^{42}\rm\,erg\ s^{-1}$ per cluster of galaxies,
\item $\sim 2 \times 10^{44}\rm\,erg\ s^{-1}$ per active galaxy, or
\item $\sim 2 \times 10^{52}$\,erg per cosmological gamma ray burst.
\end{itemize}
The coincidence between these numbers and the observed electromagnetic energy output of these sources explains why they have emerged as the leading candidates for the cosmic ray accelerators. The coincidence is consistent with the relationship between cosmic rays and photons built into the ``transparent" source previously introduced. The relationship can be extended to neutrinos.

Assuming the same energy density  of $\rho_E \sim 3 \times 10^{-19}\rm\,erg\ cm^{-3}$ in neutrinos with a spectrum $E_\nu dN / dE_{\nu}  \sim E^{-\gamma}\rm\, cm^{-2}\, s^{-1}\, sr^{-1}$ that continues up to a maximum energy $E_{max}$, the neutrino flux follows from
%
$ \int E_\nu dN / dE_{\nu}  =  c \rho_E / 4\pi  $.
%
For $\gamma = 1$ and $E_{max} = 10^8$\,GeV, the generic source of the highest energy cosmic rays produces 50 detected muon neutrinos per km$^2$ per year\cite{TKG}. [Here we have folded the predicted flux with the probability that the neutrino is actually detected given by\cite{PR} the ratio of the muon and neutrino interaction lengths in ice, $\lambda_\mu / \lambda_\nu$.] The number depends weakly on $E_{max}$ and $\gamma$. A similar analysis can be performed for galactic sources\cite{PR}.

As previously stated, for one interaction and only one, the neutrino flux should be reduced by a factor $\sim 4$. On the other hand,  there are more cosmic rays in the universe producing neutrinos than observed at earth because of the GZK-effect. The diffuse muon neutrino flux associated with the highest energy cosmic rays is estimated to be
%
$ {E_\nu}^2 dN / dE_{\nu}  \sim 5 \times 10^{-8}\rm\, GeV \,cm^{-2}\, s^{-1}\, sr^{-1} $,   
 %
to be compared to the sensitivity achieved with the first 3 years of the completed AMANDA detector of $10^{-7}\rm\,GeV\ cm^{-2}\,s^{-1}\,sr^{-1}$\cite{hill}. The analysis has not been completed but a limit has been published that is 5 times larger obtained with data taken with the partially deployed detector in 1997\cite{b10-diffuse}. On the other hand, after three years of operation IceCube will reach a diffuse flux limit of $E_{\nu}^2 dN / dE_{\nu} = 8.1 \times 10^{-9}\rm\,GeV \,cm^{-2}\, s^{-1}\, sr^{-1}$ or lower depending on the magnitude of the dominant high energy background from the prompt decay of atmospheric charmed particles\cite{ice3}.

\section{Neutrino Telescopes: First ``Light"}

While it has been realized for many decades that the case for neutrino astronomy is compelling, the challenge has been to develop a reliable, expandable and affordable detector technology to build the kilometer-scale telescopes required to do the science. Conceptually, the technique is simple. In the case of a high-energy muon neutrino, for instance, the neutrino interacts with a hydrogen or oxygen nucleus in deep ocean water and produces a muon travelling in nearly the same direction as the neutrino. The Cerenkov light emitted along the muon's kilometer-long trajectory is detected by strings of photomultiplier tubes deployed at depth shielded from radiation. The orientation of the Cerenkov cone reveals the neutrino direction.

The AMANDA detector, using natural 1 mile deep Antarctic ice as a Cerenkov detector, has operated for  more than 3 years in its final configuration of 680 optical modules on 19 strings. The detector is in steady operation collecting roughly four neutrinos per day using fast on-line analysis software. Its performance has been calibrated by reconstructing muons produced by atmospheric muon neutrinos\cite{nature}.

\begin{figure}[ht]
\centering\leavevmode
\epsfxsize=5in
\epsffile{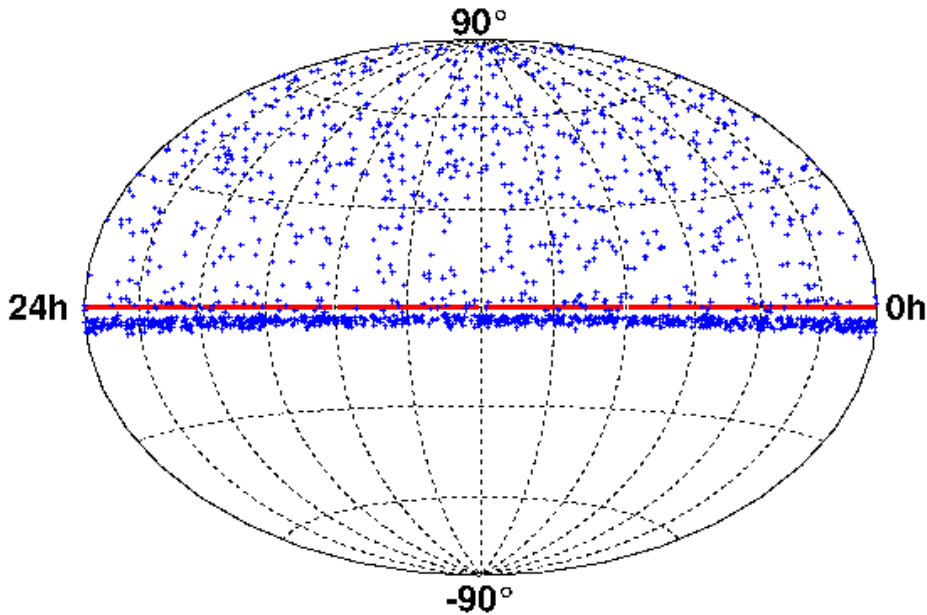}
\caption[]{Skymap showing declination and right ascension of neutrinos detected by the completed AMANDA\,II detector during its first Antarctic winter of operation in 2000.\label{fig:skyplot}}
\end{figure}

Using the first of 3 years of AMANDA\,II data, the AMANDA collaboration is performing a (blind) search for  the emission of muon neutrinos from spatially localized directions in the northern sky~\cite{HS}. Only the year 2000 data have been unblinded. The skyplot is shown in Fig.~\ref{fig:skyplot}. 90\% upper limits on the neutrino fluency of point sources is at the level of $2 \times 10^{-7}\rm\, GeV \,cm^{-2}\, s^{-1}$ or $3 \times 10^{-10}\rm \,erg\ cm^{-2}\,s^{-1}$, averaged over declination . This corresponds to a flux of $2 \times 10^{-8}\rm\, cm^{-2}\, s^{-1}$ integrated above 10\,GeV assuming a $E^{-2}$ energy spectrum typical for shock acceleration of particles in high energy sources. The most significant excess is 8 events observed on an expected background of  2.1, occurring at approximately 68 deg N dec, 21.1 hr R.A; for details see Ref.~\cite{HS}. Unblinding the data collected in 2001, 2002 may reveal sources or confirm the consistency of the year 2000 skyplot with statistical fluctuations on the atmospheric neutrino background.

With this search the AMANDA\,II detector has reached a high-energy effective telescope area of 25,000$\sim$40,000\,m$^2$, depending on declination. This represents an interesting milestone\cite{alvarezhalzen}: known TeV gamma ray sources, such as the active galaxies Markarian 501 and 421, should be observed in neutrinos if the number of gamma rays and neutrinos emitted are roughly equal as expected from cosmic ray accelerators producing pions\cite{alvarezhalzen}. Therefore AMANDA must detect the observed TeV photon sources soon, or, its observation will exclude them as the sources of the cosmic rays.

Overall, AMANDA represents a proof of concept for the kilometer-scale neutrino observatory, IceCube\cite{ice3}, now under construction. IceCube will consist of 80 kilometer-length strings, each instrumented with 60 10-inch photomultipliers spaced by 17~m.
The deepest module is 2.4~km below the surface. The strings are
arranged at the apexes of equilateral triangles 125\,m on a side. The instrumented (not effective!) detector volume is a cubic kilometer. A surface air shower detector, IceTop, consisting of 160 Auger-style Cerenkov detectors deployed over 1\,km$^{2}$ above IceCube, augments the deep-ice component by providing a tool for calibration, background rejection and air-shower physics, as illustrated in Fig.~4.

\begin{figure}[t]
\centering\leavevmode
\epsfxsize=4in
\epsffile{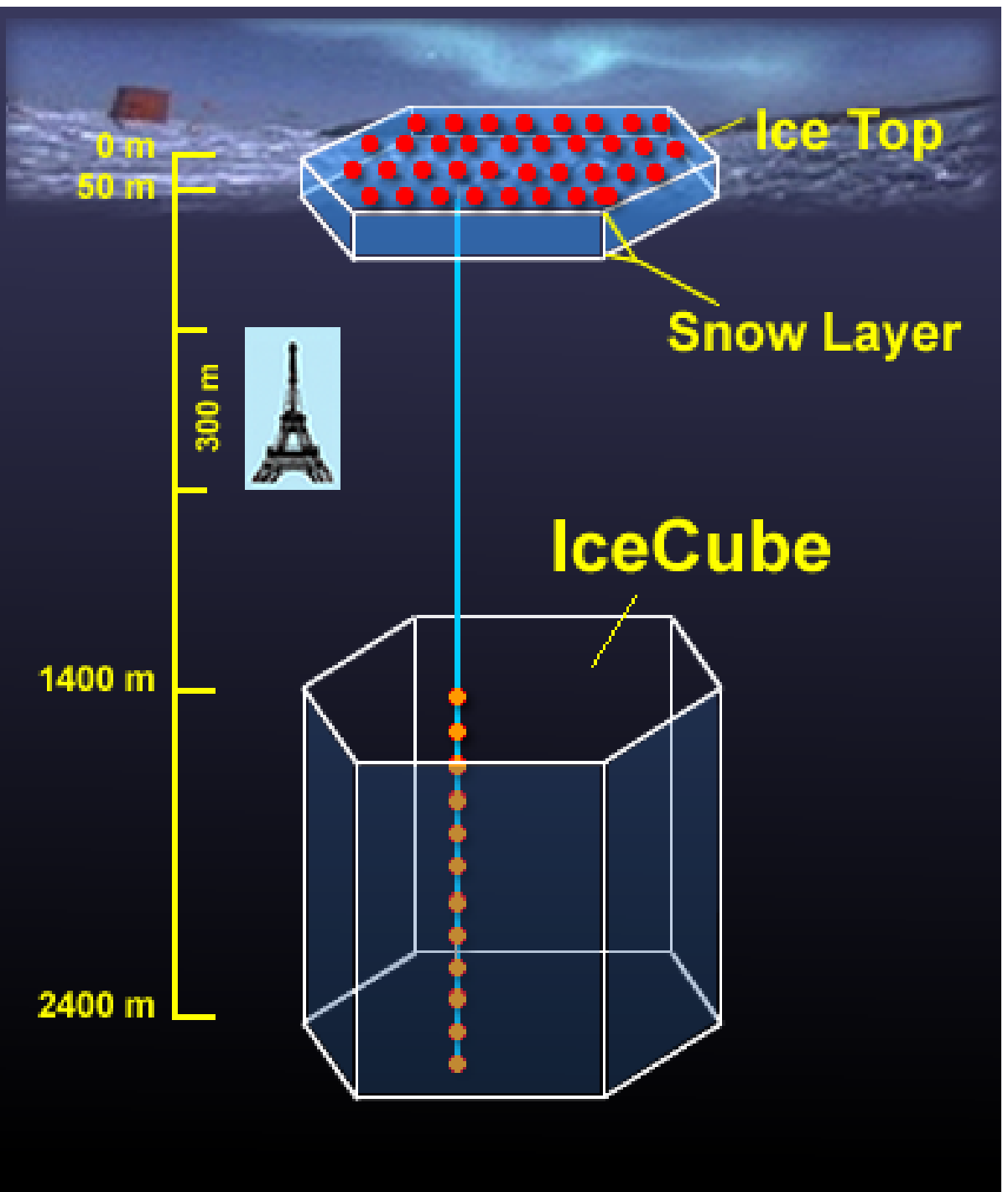}
\caption{}
\end{figure}

The transmission of analogue photomultiplier signals from the deep ice to the surface, used in AMANDA, has been abandoned. The photomultiplier signals will be captured and digitized inside the optical module.  The digitized signals are given a global time stamp with a precision of $<10$\,ns and
transmitted to the surface.  The digital messages are sent to a string
processor, a global event trigger and an event builder.  

Construction of the detector is expected to commence in the Austral summer
of 2004/2005 and continue for 6 years, possibly less.  The growing detector will take data during construction, with each string coming online within days of deployment. The data streams of IceCube, and AMANDA\,II, embedded inside IceCube,  will be merged off-line using GPS timestamps.

IceCube will offer great advantages
over AMANDA\,II beyond its larger size: it will have a higher
efficiency and superior angular resolution in reconstructing tracks, map showers from electron- and
tau-neutrinos (events where both the production and decay of a $\tau$
produced by a $\nu_{\tau}$ can be identified) and, most
importantly, measure neutrino energy. Simulations, backed by AMANDA data, indicate that the
direction of muons can be determined with sub-degree accuracy and
their energy measured to better than 30\% in the logarithm of the
energy. The direction of showers will be reconstructed to better
than 10$^\circ$ above 10\,TeV and the response in energy is linear and better than 20\%. Energy resolution is critical because,
once one establishes that the energy exceeds 1\,PeV, there is no
atmospheric muon or neutrino background in a kilometer-square detector and full sky coverage of the telescope is achieved. The background counting rate of IceCube signals is expected to be less than 0.5\,kHz per optical sensor. In this low background environment, IceCube can detect the excess of anti-$\nu_e$ events from a galactic supernova. 

\section{Mediterranean Telescopes}

Below PeV energy South Pole neutrino telescopes do not cover the Southern sky. This and the obvious need for more than one telescope anyway --- accelerator physics has clearly demonstrated the value of multiple detectors --- provide compelling arguments for deploying northern detectors. With the first observation of neutrinos by a detector in Lake Baikal with a telescope area of 2000\,m$^2$ for TeV muons\cite{baikal} and after extensive R\&D efforts by both the ANTARES\cite{antares} and NESTOR\cite{nestor} collaborations in the Mediterranean, there is optimism that the technological challenges to build neutrino telescopes in deep sea water has been met.  Both collaborations have demonstrated their capability to deploy and retrieve optical sensors, and have reconstructed down-going muons.

The ANTARES neutrino telescope is under construction at a 2400\,m deep Mediterranean site off Toulon, France. It will consist of 12 strings, each equipped with 75 optical sensors mounted in 25 triplets. The detector performance has been fully simulated\cite{antares} with the following results: a sensitivity after one year to point sources of $0.4\mbox{--}5 \times 10^{-15}\rm\, cm^{-2}\, s^{-1}$ and to a diffuse flux of $0.8 \times 10^{-7}\rm\, GeV \,cm^{-2}\, s^{-1}$ above 50\,GeV. As usual, an $E^{-2}$ spectrum has been assumed for the signal. AMANDA II data have reached similar point source limits ($0.6\mbox{--}4.2 \times 10^{-15}\rm\, cm^{-2}\, s^{-1}$) in only 196 days\cite{HS}), but it takes more than 2 years of data to reach the same diffuse limit\cite{hill}. 

Given that AMANDA and ANTARES operate at similar depths and have similar total photocathode area (AMANDA II is actually a factor of 2 smaller with 600 8\,inch versus 900 10\,inch photomultipliers) above comparison provides us with a first glimpse at the complex question regarding the relative merits of water and ice as a Cerenkov detector. The conclusion seems to be that, despite many differences, the telescope sensitivity is approximately the same for equal photocathode area. AMANDA II's performance is likely to be improved by the recently installed waveform readouts of the photomultiplier signals.

NEMO, a R\&D initiative coordinated from Catania, Sicily has been mapping Mediterranean sites and studying novel mechanical structures, data transfer systems as well as low power electronics with the goal to deploy a next-generation detector. A concept has been developed with 81 strings spaced by 140\,m. Each consists of 18 bars that are 20\,m long and spaced by 40\,m. A bar holds a pair of photomultipliers at each end, one looking down and one horizontally. First indications are that the simulated performance\cite{NEMO} is, not unexpectedly, similar to that of IceCube with a similar total photocathode area as the NEMO concept.

Recently, a wide array of projects have been initiated to detect neutrinos of the highest energies, typically above a threshold of 10 EeV, exploring other experimental signatures: horizontal air showers and acoustic or radio emission from neutrino-induced showers. Some of these experiments, such as the Radio Ice Cerenkov Experiment\cite{frichter} and an acoustic array in the Caribbean\cite{lehtinen}, have taken data; others are under construction, such as the Antarctic Impulsive Transient Antenna\cite{gorham}. The more ambitious EUSO/OWL project aims to detect the fluorescence of high energy cosmic rays and neutrinos from a detector attached to the International Space Stations. It has obtained its first levels of approval by ESA and NASA. 

For neutrino astronomy to become a viable science, several projects will have to succeed in addition to Baikal and AMANDA.
Astronomy, whether in the optical or in any other wave-band, thrives on a diversity of complementary instruments, not on ``a single best instrument".

\section*{Acknowledgments}
I thank my AMANDA/IceCube collaborators and Teresa Montaruli for discussions. This research was supported in part by the National Science Foundation under Grant No.~OPP-0236449, in part by the U.S.~Department of Energy under Grant No.~DE-FG02-95ER40896, and in part by the University of Wisconsin Research Committee with funds granted by the Wisconsin Alumni Research Foundation.

\end{document}